\documentclass[11pt,preprint]{aastex}

\newcommand\simlt{\lower.5ex\hbox{$\; \buildrel < \over \sim \;$}}
\newcommand\simgt{\lower.5ex\hbox{$\; \buildrel > \over \sim \;$}}

\begin{document}

\title{Blazar Flares from Compton Dragged Shells }
\author{Omri Golan and Amir Levinson \altaffilmark{1}}
\altaffiltext{1}{School of Physics \& Astronomy, Tel Aviv University,
Tel Aviv 69978, Israel; Levinson@wise.tau.ac.il}

\begin{abstract}
We compute the dynamics and emission of dissipative shells that are subject to a strong Compton drag, under 
simplifying assumptions about the dissipation mechanism.   We show that under conditions prevailing in blazars,
substantial deceleration is anticipated on sub-parsec and parsec scales in cases of rapid dissipation.   Such episodes
may be the origin of some of the flaring activity occasionally observed in gamma ray blazars.   The shape of the light curves
thereby produced reflects the geometry of the emitting surface if the deceleration is very rapid, or the dynamics of the shell
if the deceleration is delayed, or initially more gradual,  owing, e.g.,  to continuous injection of energy and momentum. 
\end{abstract}


\section{Introduction}

The broadband spectrum observed in blazars is dominated by beamed emission
produced in relativistic jets that emanate from the central black hole. 
These jets propagate through a dense radiative environment, and interact with seed photons 
that are supplied by extended radiation sources, notably the accretion disk around 
the black hole, gaseous clouds in the broad line region (BLR), and a dusty molecular torus located at larger scales 
(see, e.g., Joshi et al. 2014 for a recent account).
This interaction should affect the dynamics and emission of the jet.  The primary concern of most previous works
(e.g., Dermer \& Schlickeiser 1993, Sikora et al. 1994, Blandford \& Levinson 1995, Ghisellini \& Madau 1996; see also  Levinson 2006 and 
references therein)  has been the effect of this interaction on the observed spectrum, and simple emission models
that ignore dynamical effects and complex structures have been constructed for this purpose, although 
some recent works incorporate more realistic models for the dynamics of the emitting 
plasma (e.g., Joshi \& Bottcher 2011, Joshi et al. 2014, and references therein).   
In ERC models, the high-energy component of the spectral energy distribution (SED)
is attributed to inverse Compton scattering of ambient seed photons by non-thermal electrons accelerated 
in dissipative regions inside the jet.     This process, unlike synchrotron emission, gives rise to a loss of linear momentum
of emitting fluid elements and  a consequent radiative drag  that tends to decelerate the bulk flow.
In certain circumstances, explored below, this can lead to rapid, large amplitude flares. 

Rapid variability over the entire electromagnetic spectrum is a characteristic property of blazars. 
Episodic gamma-ray emission with flares durations of hours to weeks is quite typical 
to many gamma-ray blazars, with the most extreme activity recorded in the sub-class of TeV AGNs, e.g., Mrk 421, Mrk 501, PKS 2155-304.  
The observed variability time imposes a stringent constraint on the maximum size of the emission region, $\Delta r$, which in extreme
cases is inferred to be of the order of the gravitational radius of the putative black hole. 
The naive expectation is that the large amplitude short duration flares seen in gamma-ray blazars originate from small radii, as the fraction of jet energy that can be tapped for production of $\gamma$ rays scales as $\eta\simeq (\Delta r/\theta_j r_{em})^2$, where $\theta_j$ is opening angle of the jet and $r_{em}$ the emission radius, and is small for $r_{em}>>\Delta r$.
Some of the variable gamma-ray flux may be attributed to a sparking gap at the base of the jet, as proposed for M87  (Levinson 2000, Neronov \& Aharonian 2007, Levinson \& Rieger 2011), however, in powerful blazars the spectrum emitted from 
the gap is unlikely to extend beyond a few GeV, owing to the large pair-production opacity.   This, and the indications of 
correlated emission at much lower energies (radio-to-x rays)  
strongly suggest that the jets are important sources of episodic gamma-ray emission.
 It has been argued that in some cases  the observations favor models in which the variable 
gamma-ray emission originates from small regions locate at large radii (e.g., Sikora et al. 2008, Agudo et al. 2011).
Such events may be produced by a converging shock in a reconfinement nozzle (Bromberg \& Levinson 2009) 
or  magnetic reconnection in minijets (Giannois 2013). 
However, recent analysis (Nalewajko et al. 2014) challenges the far dissipation scenarios, indicating typical emission radii 
in the range 0.1-1 pc.   As shown below, on these scales the radiative drag can be substantial. 

In this paper we consider the effect of Compton drag on the dynamics and emission of dissipative 
fluid shells.  We show that if dissipation of the bulk energy commences not too far out,
the blob experiences strong deceleration that leads to large variation of the
observed flux emitted from the blob by virtue of the change in its bulk Lorentz factor.  The effect of Compton drag on the dynamics 
of a relativistic jet has been considered earlier by several authors under different assumptions (e.g., Phinney 1987, Li et al. 1992, Sikora et al. 1996).  Our method is similar to that presented in Sikora et al. (1996), however, we focus on short events that may lead to 
rapid flares, and compute the Lorentz factor profiles and the resulting gamma-ray lightcurves for a range of conditions. 
For typical ambient luminosities, the duration of flares produced by this mechanism is on the order of the 
characteristic size of the emitting blob.   Thus, flare durations as short as the dynamical time of the central engine
can naturally be accounted for in this model, provided that the bulk energy can be dissipated at a large enough rate.  A preliminary 
account of the decelerating shell model is given in Levinson (2007). Here,
we present an elaborated analysis of the dynamics of the flow, and also compute the resulting light curves. 
The construction of this model was originally motivated by the apparent discrepancy inferred in TeV blazars between the relatively large 
Doppler factors, $\delta_D\sim30-50$, required to avoid strong attenuation of the VHE flux emitted from the inner regions, 
and the much lower values, $\delta_D\sim $ a few, inferred from superluminal motions and source statistics (Georganopoulos \& Kazanas 2003, Levinson 2007).     However, the question of how the Compton drag affects the dynamics and emission of
dissipative outflows is of general interest, and is relevant essentially to all blazars.

\section{The model}

In the simplified model invoked here, a blob is ejected from a central engine of size $r_s=2GM/c^2$ and accelerated to a Lorentz factor $\Gamma_0$.  When it reaches  some radius $r_d\simgt\Gamma_0^2 r_{s}\simeq10^{17}\Gamma^2_{20}M_9$ cm, dissipation suddenly starts, e.g., via formation of shocks or explosive conversion of magnetic energy, leading to continuous particle acceleration. Inverse Compton scattering of ambient photons then leads to a strong radiative drag that tends to decelerate the blob.   In certain circumstances, this drag may be compensated by injection of energy and momentum into the blob by some external 
agent, delaying the deceleration of the emitting plasma.    The deceleration may be delayed also in cases where the dissipation commences very close to the central engine, where the radiation field is dominated by direct illumination from the disk and is highly unisotropic, as in
this zone the blob may propagate at the equilibrium Lorentz factor  until reaching scales where the external radiation field is roughly isotropic (e.g., Sikora et al. 1996, Vuillaume et al. 2014).   In the present model we shall refer to this case as delayed deceleration.    
In reality, the structure and velocity of the emitting plasma are expected to be non-uniform by virtue of confined dissipation and rapid cooling. 
In what follows we ignore such complications and compute the dynamics and
emission of the blob under simplifying assumptions about the microphysics of dissipation. 

The energy and momentum  fluxes of the emitting fluid are given by,
\begin{eqnarray}
& & T^{0r}=(w+B^{ 2}/4\pi)\Gamma^2\beta_\Gamma \equiv  u_j^\prime\Gamma^2\beta_\Gamma,\label{T0r}\\
& & T^{rr}=(w+B^{ 2}/4\pi)\Gamma^2\beta^2_\Gamma+p+B^2/8\pi
\label{Trr}
\end{eqnarray}
where $w=\rho+e+p$ is the proper specific enthalpy, $p$, $\rho$ and $e$ are the pressure, proper density 
and proper internal energy, respectively, $B$ is the proper magnetic field, and $\Gamma=(1-\beta_\Gamma^2)^{-1/2}$ is the bulk Lorentz factor.   In terms of the total 
power,  $P_j=cT^{0r}\pi \theta_j^2 r^2$, where
$\theta_j$ is the half opening angle of the flow, we have
\begin{equation}
u'_j=\frac{P_j}{\pi \theta_j^2 r^2 c\Gamma^2}=0.1 P_{j44}(\theta_j\Gamma r_{17})^{-2}\quad \rm erg\ cm^{-3},
\end{equation}
for $\beta_\Gamma=1$,  $r=10^{17}r_{17}$ cm and $P_j=10^{44}P_{j44}$ erg s$^{-1}$.

For simplicity, we assume that the ambient radiation field is roughly isotropic in the frame of the central engine, and has a luminosity $L_s$. 
This quasi-isotropic radiation field is contributed by scattering and reprocessing of the central 
UV radiation by  gas in the broad line region, and by emission from a dusty torus (e.g., Joshi et al. 2014).   
Here, $L_s$ represents the sum of these components, and is a fraction of the total luminosity of the continuum source.
The radial  profile of the ambient intensity intercepted by the jet depends on the geometries of the broad line region
and the dusty torus (e.g., Joshi et al., 2014).   
Denoting $x=r/r_d$, we express it as
\begin{equation}
I_s(\epsilon_s,\mu,r)=\kappa_s(r)F(\epsilon_s),\quad \kappa_s(r)=\frac{L_s}{16\pi^2r_d^2c}f_s(x),\label{Is-star}
\end{equation}
with the normalization $\int_0^\infty F(\epsilon_s)d\epsilon_s=1$,  such that the total energy density
at radius $r$ is $u_s(r)=4\pi\kappa_s(r)$.    For central emission, $f_s(x)=x^{-2}=(r_d/r)^2$.    
For a flat profile within some scale $r_l > r_d$, $f_s(x)=(r_d/r_l)^2$ at $x<r_l/r_d$. 
For the spectrum of the ambient radiation field we adopt 
\begin{equation}
F(\epsilon_s)\propto \left\{\begin{array}{lll}
          \epsilon_s^{1/2}
 & \mbox{\ $10^{-8} <\epsilon_s/m_ec^2<10^{-4}$},\\
        \epsilon_s^{-3/2} & \mbox{\ $10^{-4} <\epsilon_s/m_ec^2<10^{-1}$},\\
0 &  \mbox{\ else}\end{array} \right. 
\label{h_char-tot}
\end{equation}
that mimics a typical soft spectrum.  

In the rest  frame of the blob the intensity is given by 
\begin{equation}
I^\prime_s(\epsilon^\prime_s,\mu^\prime,r)=\frac{\kappa_s(r)}{\Gamma^3(1+\beta_\Gamma\mu^\prime)^3}F(\epsilon^\prime_s\Gamma(1+\beta_\Gamma\mu^\prime)),\label{Is-exact}
\end{equation}
and the comoving energy density by
\begin{equation}
u^\prime_s=2\pi\int I^\prime_s d\epsilon^\prime_s d\mu^\prime=2\pi\kappa_s(r)\int_{-1}^1{\frac{d\mu^\prime}{\Gamma^4(1+\beta_\Gamma\mu^\prime)^4}}=u_s\Gamma^2(1+\beta_\Gamma^2/3).\label{us-exact}
\end{equation}
From Equation (\ref{Is-exact}) it is seen that the comoving intensity peaks sharply around the direction opposite to the blob velocity, viz., $\mu'=-1$.
To simplify our calculations we approximate the comoving intensity as a beam moving in the direction $\mu'=-1$:
\begin{equation}
I^\prime_s(\epsilon^\prime_s,\mu^\prime,r)=\eta_s(r)F(\epsilon_s^\prime\Gamma(1+\beta_\Gamma\mu^\prime))\delta(1+\mu^\prime),\label{I-beamed}
\end{equation}
with $\eta_s(r)=\kappa_s(r)2\Gamma(1+\beta_\Gamma)^{-1}(1+\beta_\Gamma^2/3)\simeq4\Gamma\kappa_s(r)/3$.  
It can be readily verified that with this choice the comoving energy density satisfies Equation (\ref{us-exact}). 

We further suppose that the electron distribution function is isotropic in the fluid rest frame and can be approximated as a power
law: $dn^\prime_e/d\gamma=\kappa_e(r) \gamma^{-q}$;  $\gamma_{1}<\gamma<\gamma_{2}$, where $m_ec^2 \gamma$ 
is the corresponding electron energy, as measured in the comoving frame.  We define $\xi_e(r)$  to be the fraction of 
the total internal energy carried by
the relativistic electrons, that is, $\xi_e e=u'_e$, where $u'_e=\int m_ec^2\gamma dn'_e$, subject to the boundary condition 
$\xi_e(r\le r_d)=0$.  In terms of this parameter 
$\kappa_e=\xi_e(u'_j/m_ec^2)(2-q)/(\gamma_{2}^{2-q}-\gamma_{1}^{2-q})$, and we note that $ (\gamma_{2}^{2-q}-\gamma_{1}^{2-q})/(2-q)=\ln(\gamma_{2}/\gamma_{1})$ in the limit  $q\rightarrow 2$.    We suppose that $\xi_e$ reaches a maximum value over some characteristic length scale that depends on the microphysics of the specific dissipation mechanism.   This length scale is typically of the order of  the 
gyroradius of accelerated electrons, which is shorter than the cooling time of the highest energy electrons.  Thus, we take $\xi_e$ to be 
constant at $r>r_d$.

The maximum Lorentz factor of the electron distribution, $\gamma_{2}$, is likely to be limited by cooling.  In the  limit of  Bohm diffusion the 
acceleration time is $t^\prime_{acc}=(\gamma m_ec)/(\eta_{acc}eB)$.  Equating with the Compton cooling time, $t_c^\prime=\gamma m_ec^2/P_{com}=m_e c/(\gamma \sigma_T u_s^\prime)$,   and adopting $\eta_{acc}=0.1$, we estimate 

\begin{equation}
\Gamma\gamma_{2}\simeq3\times 10^7(B/B_0)^{1/2}\xi^{1/4}_{B-1}P_{j044}^{1/4}L_{s44}^{-1/2}(\Gamma_0\theta_j)^{-1/2}r^{1/2}_{d17}f_s^{-1/2}
\label{gmax}
\end{equation} 
in terms of the fraction $\xi_B=0.1\xi_{B-1}=u'_B/u'_j$, $u'_B=B^2/8\pi$.  Here $B_0=B(r=r_d)$ is the magnetic field at the onset of dissipation.

\subsection{Dynamics}
We consider only situations in which dissipation is rapid enough to produce high enthalpy inside the blob on a timescale comparable to the radiative time.  Then, one can adopt a simple prescription in which the dissipation is formally 
treated as an initial condition (at $r=r_d$) of the flow equations.  The magnetization of the flow is then expected 
to be low, so that the effect of the magnetic field on the dynamics of the system can be neglected. 
This may not apply to cases were the dissipation time is much longer than the light crossing time of the shell.
We suppose that the dissipated energy is redistributed in a way that a fraction $\xi_e$ of the total dissipation energy is injected in the form of a power law electron 
distribution, as explained above.  Under these assumptions, the dynamics of the blob subsequent to the onset of dissipation is governed by the equation
\begin{equation}
\partial_\mu T^{\mu\nu}=S_c^{\nu},
\label{eq:mot-gen}
\end{equation}
where the source terms $S_c^\nu$ account for the radiative drag acting on the blob, and are given explicitly in Equations (\ref{S_c^0-lab})- (\ref{S_c^i-lab}).  
For a conical expansion, the $0$ and $r$ components of Equation (\ref{eq:mot-gen}) can then be written, to order $O(\Gamma^{-2})$, in the form (see appendix \ref{app:Eq-mot})
\begin{eqnarray}
w\frac{d}{dt}\ln \Gamma+\frac{dp}{dt}-\Gamma^{-2}\frac{\partial p}{\partial \zeta} =S_c^r-\beta_\Gamma S_c^0,\label{mot-radial-0}\\
\frac{1}{r_2^2}\frac{d}{dt}(r_2^2w\Gamma^2)-\frac{d}{dt}p-\frac{\partial p}{\partial \zeta} -w\frac{\partial}{\partial \zeta} \ln\Gamma=S_c^0,
\label{mot-radial-r}
\end{eqnarray}
in terms of the coordinates $t$ and $\zeta(t,r)=r_2(t)-r$, where $d/dt=\partial_t+\beta_\Gamma\partial_r$ is the Lagrangian derivative,
and $r_2(t)=ct+O(\Gamma^{-2})$ denotes the trajectory of the blob's front.   The interior of the blob encompasses the range 
$0\le \zeta \le \Delta$.

Now, for the relativistic electron population invoked above $<\gamma^2\beta^2>=<\gamma^2> >>1$,  where 
\begin{equation}
<\gamma^2>=\frac{1}{n'_e}\int {\gamma^2 } {\frac{dn^\prime_e}{d\gamma} d\gamma},\quad n^\prime_e=\int_{\gamma_{min}}^{\gamma_{max}} {\frac{dn^\prime_e}{d\gamma} d\gamma},
\end{equation}
and to order $O(\Gamma^{-2})$ the source terms, Equations (\ref{S_c^0-lab}) - (\ref{S_c^i-lab}), are given by 
\begin{equation}
S^0_c=-\frac{8}{3}\Gamma^3<\gamma^2>u_s\sigma_T n^\prime_{e},
\label{S0c}
\end{equation}
and 
\begin{equation}
S_c^r=\beta_\Gamma S_c^{0}+S_c^0/3\Gamma^2.  
\end{equation}

\subsubsection{Uniform blob}

We consider first a uniform blob ($\partial_\zeta=0$) having a fixed length $\Delta$ in the frame of the central engine.  
Its motion is then characterized by a single Lorentz factor at all times,  $\Gamma(\zeta,t)=\Gamma(t)$.
We suppose that microphysical processes redistribute the dissipation energy in a way that a fraction $\xi_e$ of the 
total internal energy is uniformly injected in the form of a power law electron distribution inside the blob.  For clarity, we
ignore here the contribution of the thermal population to the Compton drag.  As shown below, this is justified when $\xi_e\chi\gamma_{2}m_ec^2$ is larger than the thermal energy, here $\gamma_{2}$ is the upper cutoff of the electron distribution given in 
equation (\ref{gmax}), and $\chi$ is defined below.    The inclusion of the thermal electrons in the source terms does not alter our results significantly, and at any rate, if the dissipation produces a population of relativistically hot electrons with a thermal 
energy $\gamma_Tm_ec^2$ rather than a power law distribution, 
it can be readily  accounted for by taking $\chi\xi_e=1$, $\gamma_2=\gamma_T$ in Equation (\ref{alpha}) below.

We find it convenient to use the parametrization $<\gamma^2>/<\gamma>=\chi\gamma_{2}$, where 
$<\gamma>m_ec^2=u^\prime_e/n'_e$ is the average energy of nonthermal electrons. 
For the power law energy distribution invoked above we have 
\begin{equation}
\chi=\frac{(2-q)}{(3-q)}\frac{[1-(\gamma_{2}/\gamma_{1})^{3-q}]}{[1-(\gamma_{2}/\gamma_{1})^{2-q}]}\frac{\gamma_1}{\gamma_2}.
\label{chi}
\end{equation}
With $\gamma_2>>\gamma_1$ and $q<2$ it gives $\chi\simeq (2-q)/(3-q)$.  For $q=2$ we have 
$\chi=[\ln (\gamma_2/\gamma_1)]^{-1}>0.1$, and for $q=2.5$, $\chi\simeq \sqrt{\gamma_1/\gamma_2}$.  A reasonable choice for
the minimum electron energy, adopted henceforth, is $\gamma_1=m_p/m_e$.
We further define the fiducial coordinate $x=ct/r_d$, and write $u_s=u_{s0}f_s(x)/f_{s0}$, denoting
$f_{s0}=f_s(x=1)$.   Equations (\ref{mot-radial-0}) and (\ref{mot-radial-r}) 
with $\partial_\zeta p=\partial_\zeta \Gamma=0$ can then be re-expressed as 
\begin{eqnarray}
&&\frac{d }{d x}\ln(\Gamma^2x^2w)=-6\alpha(\chi/\chi_0)(\gamma_2/\gamma_{20}) (\Gamma/\Gamma_0)(f_s/f_{s0})(4e/3w),
\label{eq-mot-v2}\\
&&d\ln\left(\frac{wx^2}{\Gamma}\right)-\frac{3p}{w}d\ln p=0,\label{eq-state-v2}
\end{eqnarray}
in terms of the constant 
\begin{equation}
\alpha=\frac{r_d\sigma_T\chi_0\xi_{e}\Gamma_0\gamma_{\rm 20}u_{s0}}{3m_ec^2}=
3\times10^{4}\chi_0\xi_{e}\xi^{1/4}_{B-1}P_{j044}^{1/4}L_{s44}^{1/2}r_{d17}^{-1/2}f_{s0}^{1/2}.
\label{alpha}
\end{equation}
The solution of these coupled equations depend on the equation of state and the assumptions about $\gamma_2$.   Approximate analytic solutions can be obtained in the case $q\le2$ for which $\chi\simeq \chi_0$ is a good approximation. For a relativistically hot blob, we adopt the equation of state $e=3p$.  We then obtain 
\begin{eqnarray}
&&\frac{d }{d x}\ln(\Gamma/x)=-\alpha(\gamma_2/\gamma_{20}) (\Gamma/\Gamma_0)(f_s/f_{s0}).
\label{eq-mot-v4}
\end{eqnarray}
As a first example, let us take the maximum electron energy to be constant during the deceleration phase, that is., 
$\gamma_2=\gamma_{20}$.  The solution then reads:
\begin{equation}
\Gamma(x)=\frac{\Gamma_{0}x}{[1+\alpha g_1(x)]},\label{G-sol1}
\end{equation}
where $g_1(x)=\int_1^x [x^\prime f_s(x^\prime)/f_{s0}]dx^\prime$, with $g_1(x)=\ln x$ for $f_s=x^{-2}$ and $g_1(x)=x^2/2$ for a
flat profile, $f_s/f_{s0}=1$.   As a second example, we suppose that $\gamma_2$ is given by Equation (\ref{gmax}), 
and that the magnetic field evolution is dictated by the ideal MHD limit, viz., $B/B_0=\Gamma_0/(x \Gamma)$.   
We then obtain the solution
\begin{equation}
\Gamma(x)=\Gamma_{0}{[1-\alpha g_2(x)]^2},\label{G-sol2}
\end{equation}
with $g_2(x)=\int_1^x x^{\prime-1} [f_s(x^\prime)/f_{s0}]^{1/2}dx^\prime$.
It is worth noting that these analytic solutions hold only for times $x$ at which the blob is relativistically hot, and $\Gamma(x)>>1$. 

\begin{figure*}
\centerline{\includegraphics[scale=0.8]{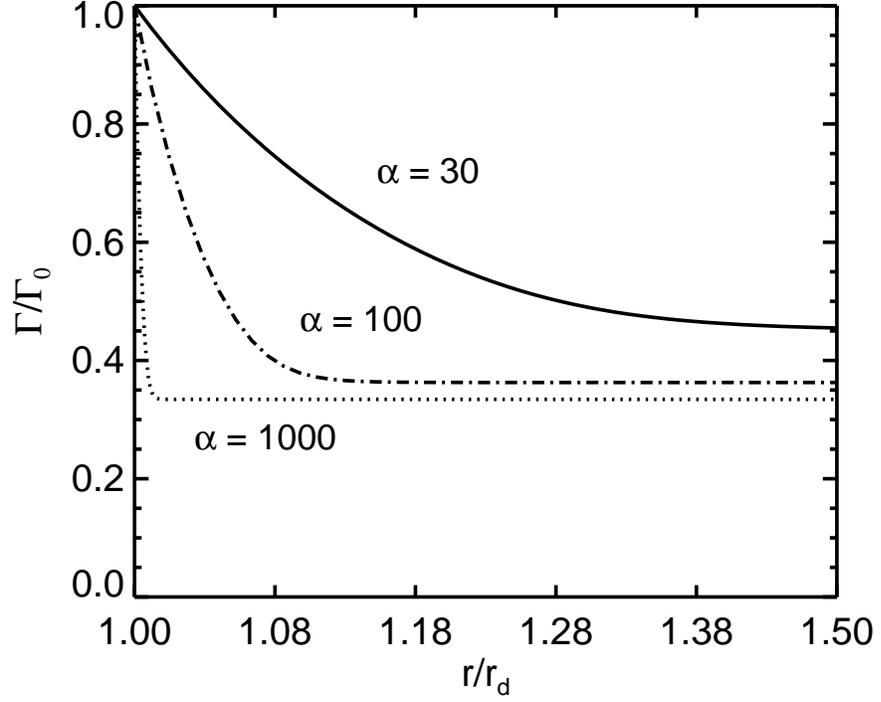}}
\caption{Lorentz factor profiles computed numerically using 
Equations (\ref{eq-mot-v2})-(\ref{eq-state-v2}) with cooling limited injection
(Eq. \ref{gmax}), and different values of $\alpha$, as indicated. }
\label{fig-f1}
\end{figure*}

Figure \ref{fig-f1}  exhibits exact numerical solutions of Equations (\ref{eq-mot-v2})-(\ref{eq-state-v2}) for $\rho_0c^2/p_0=10^{-2}$,
 $\gamma_2$ given by Equation (\ref{gmax}) with $B/B_0= \Gamma_0/(x\Gamma)$,  intensity profile $f_s=x^{-2}$, 
 and different choices of $\alpha$.   As a check, we obtained solutions also for different intensity profiles 
and different assumptions about $\gamma_2$,  and found little differences in the Lorentz factor profiles 
for a given $\alpha$ in the regime of rapid deceleration.   The key parameter that determines the dynamics is $\alpha$. 

\subsubsection{\label{sec:shocks} Internal shocks}
The model outlined above assumes uniform acceleration of electrons to nonthermal energies at all times.  This assumption 
may hold in certain situations but not in general.     In shocks, for instance,  particle acceleration is confined to a region around the 
shock front of a characteristic size comparable to the gyroradius of the accelerated electrons, as measured in the shock frame.  
For the highest energy electrons it is roughly equal to the cooling distance (if not limited by escape).    For lower energy electrons
it may be even smaller.   As the downstream fluid 
moves away from the shock, electrons of energy $\gamma m_ec^2$ cool over time 
$t_c^\prime=m_e c/(\gamma \sigma_T u_s^\prime)$, where $u_s^\prime=4\Gamma^2u_s/3$ is the comoving energy density 
of the external radiation (see Eq. \ref{us-exact}).  The distance traversed by these electrons in the black hole frame 
before cooling down is  $l_{c}=\Gamma t^\prime_{c} c=3 m_e c^2/(4\Gamma \gamma \sigma_T u_s)$.    
Using Equation (\ref{alpha})  then yields  $\alpha l_c/r_d=\chi\xi_e<< 1$ for $\xi_e<<1$.    Consequently, IC scattering off nonthermal electrons 
should not lead to significant deceleration of the blob, unless the entire shock energy is converted to non-thermal electrons ($\chi\xi_e=1$).  However, about half the shock energy is carried by a population 
of thermal electrons having an average comoving energy $\gamma_{T0}m_ec^2\simeq \Gamma_{sh}m_pc^2/2$ just 
downstream of the shock, where $\Gamma_{sh}$ denotes the shock Lorentz factor
(that is, the Lorentz factor of the upstream fluid measured in the shock frame).    These electrons cool as they propagate away from the shock.   The Lagrangian rate of change of the
thermal energy in the downstream flow is governed by the equation
\begin{equation}
\frac{d\gamma_T}{d t}=-\frac{\gamma_T}{\Gamma t^\prime_c }=-\frac{4\sigma_Tu_s}{3m_ec}\Gamma\gamma_T^2.\label{gamm_T-cool}
\end{equation} 
To compute the shock structure one needs to solve Equations (\ref{mot-radial-0})-(\ref{mot-radial-r}) coupled to Equation (\ref{gamm_T-cool}).  
Such treatment is beyond the scope of this paper.  To illustrate the effect of radiative drag on the shock we invoke the uniform 
shell approximation, that is, keep only the Lagrangian derivatives in Equations (\ref{mot-radial-0})-(\ref{mot-radial-r}). 
Then, adopting $<\gamma^2>m_ec^2n^\prime_e/w=4\gamma_T$ and $w=4p$ we obtain 
the rate of  change of the bulk Lorentz factor:

\begin{equation}
\frac{d}{d t}\ln(\Gamma/x)=-\frac{16\sigma_Tu_s}{9m_ec}\Gamma\gamma_T.\label{Gamm-shock}
\end{equation} 
With $u_s=u_{s0}f_s(x)/f_{s0}$, the solution of the coupled Equations (\ref{gamm_T-cool}) and (\ref{Gamm-shock}) reads:
\begin{equation}
\Gamma(x)=\Gamma_0\frac{x}{[1+\alpha_Tg_T(x)]^{4/7}},\label{G-sol-Therm}
\end{equation}
here $g_T(x)=\int_1^x x^{\prime 7/4}[f_s(x^\prime)/f_{s0}]dx^\prime$, and 
\begin{equation}
\alpha_T=\frac{28r_d\sigma_T\Gamma_0\gamma_{T0}u_{s0}}{9m_ec^2}\simeq4\Gamma_0\Gamma_{sh}L_{s44}r_{d17}^{-1}f_{s0}.\label{alphaT}
\end{equation} 
Since $\Gamma_0\Gamma_{sh}>>1$ significant deceleration is expected on sub-parsec scales.  For instance, assuming a flat 
intensity profile below the radius $r_l=10^{18}$ cm, with $u_{s0}=10^{-3}$ erg cm$^{-3}$ at $r<r_l$, 
we obtain $\alpha_T=0.2\Gamma_0\Gamma_{sh}r_{d17}$.  Thus, colliding shells having a Lorentz factor $\Gamma_0>10$ will experience 
substantial deceleration. 

Now, the deceleration of the downstream plasma should lead to a gradual strengthening of the reverse shock and a weakening 
of the forward shock.  Our preliminary calculations indicate substantial over-compression of the reverse shock already at $\alpha_T\sim$ a 
few.  In the frame of the central engine this translates to a deceleration of the entire shocked shell, as 
described qualitatively by the simple blob model outlined in the preceding section.     The strengthening of the reverse shock should lead
to enhanced dissipation rate, whereby the bulk energy of the unshocked shell is ultimately radiated away with high efficiency.  
Consequently, as long as the beaming cone of emission is narrower than the angular extent of the shell, the total flux observed will remain roughly constant.   Thus, the delayed-deceleration  model is relevant for the evolution of the total flux.  However, the change in the Doppler factor resulting from the deceleration of the shocked shell leads to a change in the observed energy of scattered photons and this, in turn,  can significantly alter the  evolution of the SED.   In particular, we anticipate 
different durations and times of peak emission of flares observed in different energy bands.   The non-uniformity of the emitting
plasma downstream of the shock adds complexity.   

A comprehensive analysis of emission from internal shocks in blazars is given in Joshi \& Bottcher (2011) and Joshi et al. (2014)  
ignoring the effect of Compton drag on the dynamics of the shock.   The decay of the emission in
their model is  due to cooling of the emitting electrons following shock crossing.  This situation is well represented by 
the delayed deceleration  model we adopt below.   The neglect of Compton
drag is justified only at radii where $\alpha_T<1$ in Equation (\ref{alphaT}).   As explained above, the inclusion of 
Compton drag and non-uniform particle acceleration should have a profound effect on the light curves.  A complete treatment of the 
dynamics and emission of internal shocks that are subject to a strong Compton drag will be presented in a future publication.

\subsection{Inverse Compton emission}

As will be shown below, the shape of the lightcurves reflects the geometry and dynamics of the emitting material.   In particular, 
time retardation associated with the curvature of the emitting surface sets a limit on the rise and decay times of the flare.   Furtheremore,
as mentioned above certain situations can be described by  delayed deceleration of the blob, that can have an important effect.  To
illustrate such effects, we shall consider also cases in which the  Lorentz factor remains constant at its initial value $\Gamma_0$ up to some
radius $r_d< r_{dec}< r_d+\Delta/(1-\beta_\Gamma)$, and only then deceleration commences.   In reality, the structure of the
emitting zone is expected to be non-uniform in those cases,  depending on the specific model.  Those details may affect the resulting
emission.   Our purpose here is merely to illustrate how dynamical effects are imprinted in the lighcurves.   For this purpose our 
simple treatment of delayed deceleration is sufficient.     Moreover, this prescription also describes situations in which the radiative 
drag is too small to affect the dynamics of the shell, and the decay of the emission at the end of the dissipation phase is due to cooling of the emitting electrons. 

Since we are mainly interested here in the high-energy emission from the blob, we consider only the contribution of Inverse Compton scattering. 
The spectral evolution measured by a distant observer is computed as follows:  For a given choice of parameters we first solve  Equations (\ref{eq-mot-v2})-(\ref{eq-state-v2})
numerically to obtain  the  Lorentz factor profile  $\Gamma(r)$ in the frame of the black hole.  Adopting the beam approximation, the intensity of the background  radiation at  any  radius $r$ is then transformed into the rest frame of blob using $\Gamma(r)$ in Equation (\ref{I-beamed}).  
The comoving intensity  thereby obtained is used to compute the comoving emissivity of the scattered radiation, 
$j^\prime_{sc}(\epsilon^\prime,\mu^\prime,r)$.  The emissivity in the black hole frame is given by
$j_{sc}(\epsilon, \mu,r)=[\Gamma(1-\beta_\Gamma\mu)]^{-2} j_{sc}^\prime(\epsilon^\prime,\mu^\prime,r)$,
where $\mu=(\mu^\prime+\beta_\Gamma)/(1+\beta_\Gamma\mu^\prime)$ and $\epsilon=\Gamma(1+\beta_\Gamma\mu^\prime)\epsilon^\prime $.
The scattered intensity emitted by the blob is obtained upon integrating the emissivity across the blob, taking into account the time delay between 
emission of photons from different locations:
\begin{equation}
I_{sc}(\epsilon,\mu,r)=\int_0^{\Delta r}j_{sc}(\epsilon,\mu,r-y)dy \label{I-obs}
\end{equation}
where the distance $\Delta r$ is related to the blob's length $\Delta$ through
\begin{equation}
\Delta=\int_{r-\Delta r}^{r}[1-\beta_\Gamma(r^\prime)\mu]dr^\prime,
\end{equation}
and is a function of $r$. In deriving Equation (\ref{I-obs}) we assumed that the emissivity is uniform 
inside the blob and vanishes outside it.  Note that
 \begin{equation}
 t_{ob}(r)=\int_{r_d}^r[1-\beta_\Gamma(r^\prime)\mu]dr^\prime/c
 \label{t_ob}
\end{equation}
is the time measured by a distant observer viewing the blob at an angle $\theta=\arccos(\mu)$ relative to its direction of 
motion, so that formally $\Delta/c=t_{ob}(r)-t_{ob}(r-\Delta r)$.

\begin{figure*}
\centerline{\includegraphics[scale=0.8]{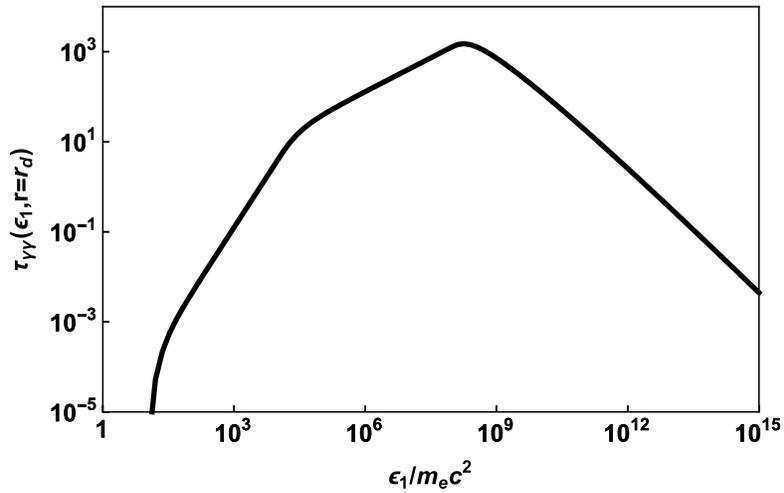}}
\caption{Pair-production optical depth as a function of photon energy .}
\label{fig-f2}
\end{figure*}

High energy photons emitted by the blob will be attenuated by pair production on background photons.    Here we model this attenuation by
an exponential cutoff at the corresponding pair production optical depth $\tau_{\gamma\gamma}(\epsilon,r)$.  The latter is computed using Equations (3.1)-(3.3) in Blandford \& Levinson (1995), and plotted in figure \ref{fig-f2}. 
The spectral flux measured by a distant observer viewing the blob at time $t_{obs}$ at an angle $\theta$ satisfies
\begin{equation}
 {\cal F}^\infty (\epsilon,\mu,t_{ob}) \propto \int_{0}^{y_0} dy\int_{\Sigma(r-y)}j_{sc}(\epsilon, \mu,r-y)e^{-\tau_{\gamma\gamma}(\epsilon,r-y)}d\Sigma^\prime,
 \label{ob-flux}
\end{equation}
where the radius $r$ is computed at the observed time $t_{ob}$ using Equation (\ref{t_ob}), that is $r=r(t_{ob})$, and the inner integration is over the emitting surface of the blob at the retarded time $(r-y)/c$.    

In the immediate  deceleration case, the rise and decay times of the observed flux are dominated by temporal delays associated with 
the curvature of the emitting surface.   For a conically expanding shell the emitting surface is spherical, and Equation (\ref{ob-flux}) reduces to:
\begin{equation}
 {\cal F}^\infty (\epsilon,\mu,t_{ob}) \propto \int_{0}^{y_0} dy\int_{\mu_1(r,y)}^{\mu_2(r,y)}j_{sc}(\epsilon, \mu,r-y)(r-y)^2e^{-\tau_{\gamma\gamma}(\epsilon,r-y)}\mu^{-3}d\mu,
 \label{ob-flux-curvt}
\end{equation}
with
\begin{eqnarray}
& &\mu_2(r,y)=\max \left\{\cos\theta_j, 1-\frac{1}{r-y}\int_{r-y}^r[\beta^{-1}_\Gamma(y^\prime)-1]dy^\prime\right\},\\
& &\mu_1(r,y)=\min \left\{1,1-\frac{(1-\mu_2)(r-y)-\Delta}{r-y-\Delta}\right\},
\end{eqnarray}
where $\theta_j$ is the opening angle of the flow, and $y_0$ is determined from the condition $\mu_1(r,y_0)=\cos\theta_j$.

\begin{figure*}
\centerline{\includegraphics[scale=0.6]{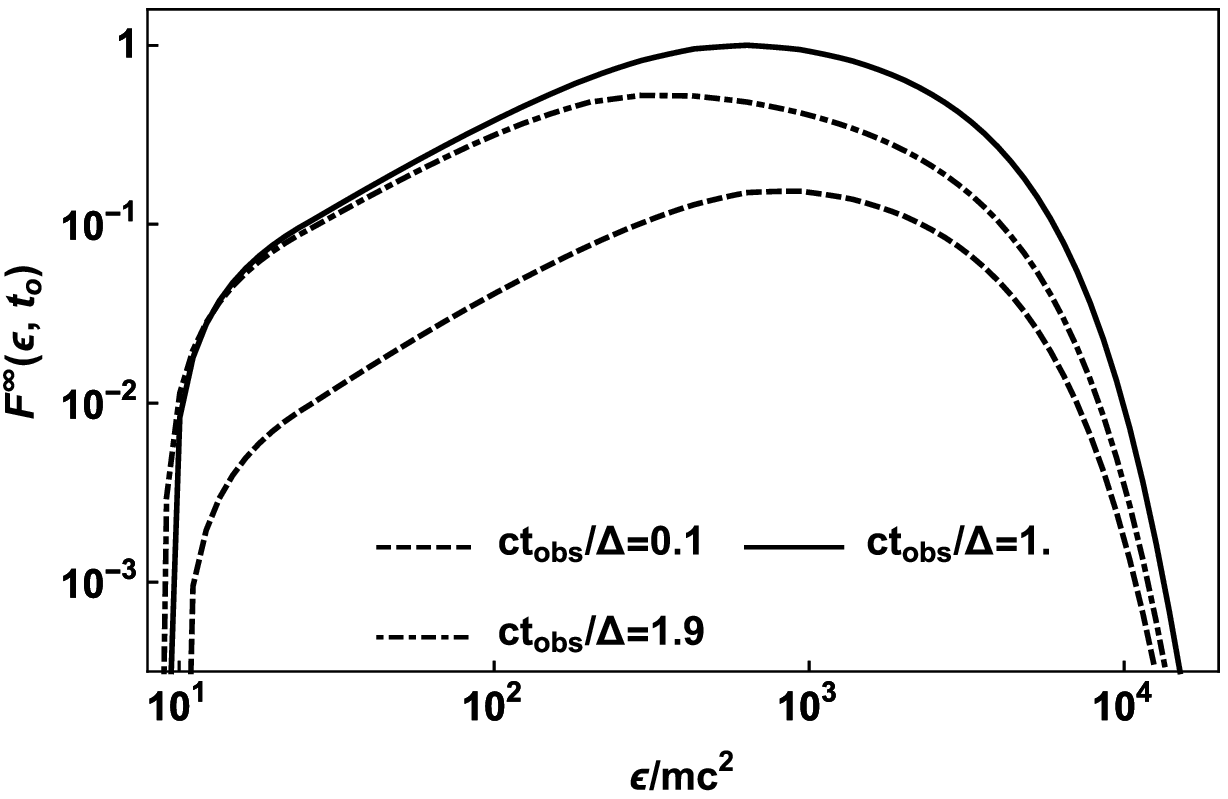}\includegraphics[scale=0.6]{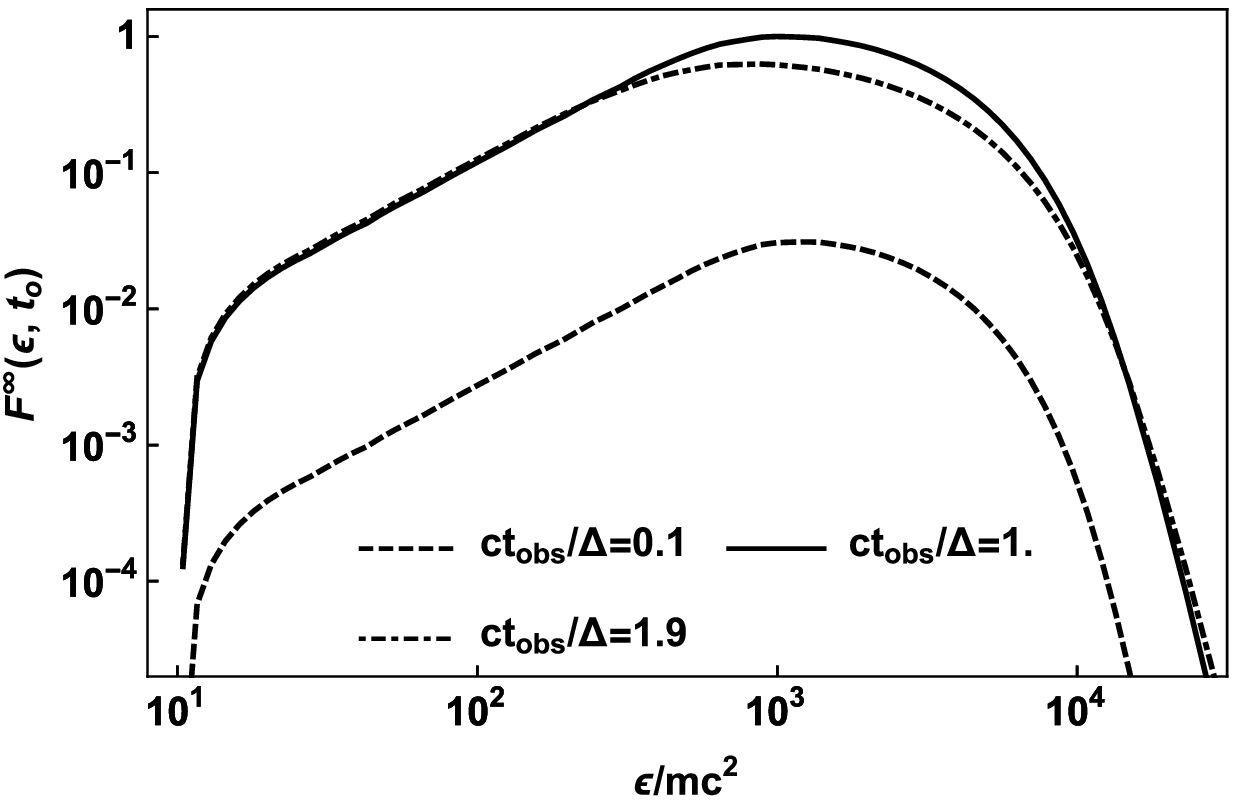}}
\caption{Spectral energy distribution of ERC radiation emitted from a decelerating blob following onset of dissipation, for 
a non-delayed deceleration (left panel) and delayed deceleration with $r_{dec}-r_d=\Delta/(1-\beta_\Gamma)$ (right panel). }
\label{fig-f3}
\end{figure*}

\begin{figure*}
\centerline{\includegraphics[scale=0.8]{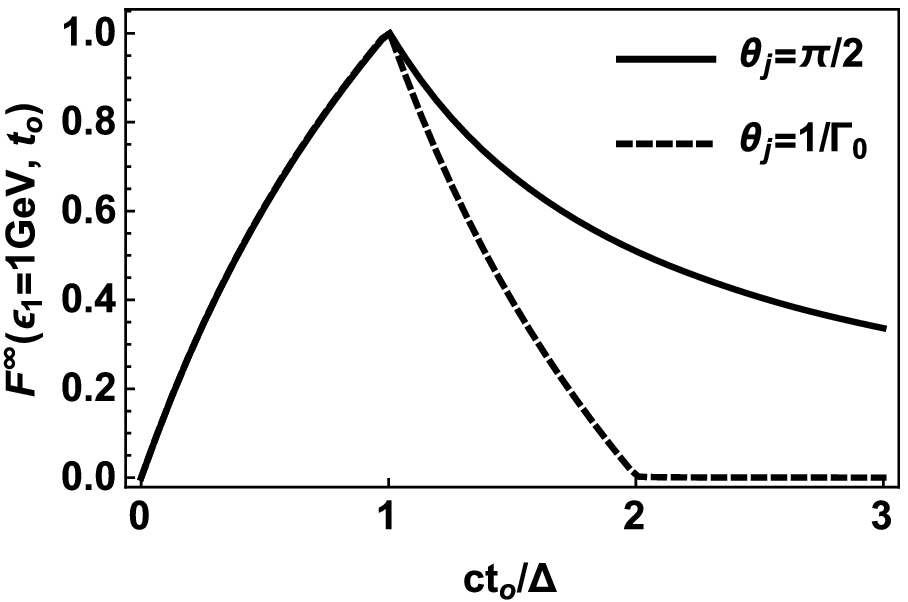}\includegraphics[scale=0.8]{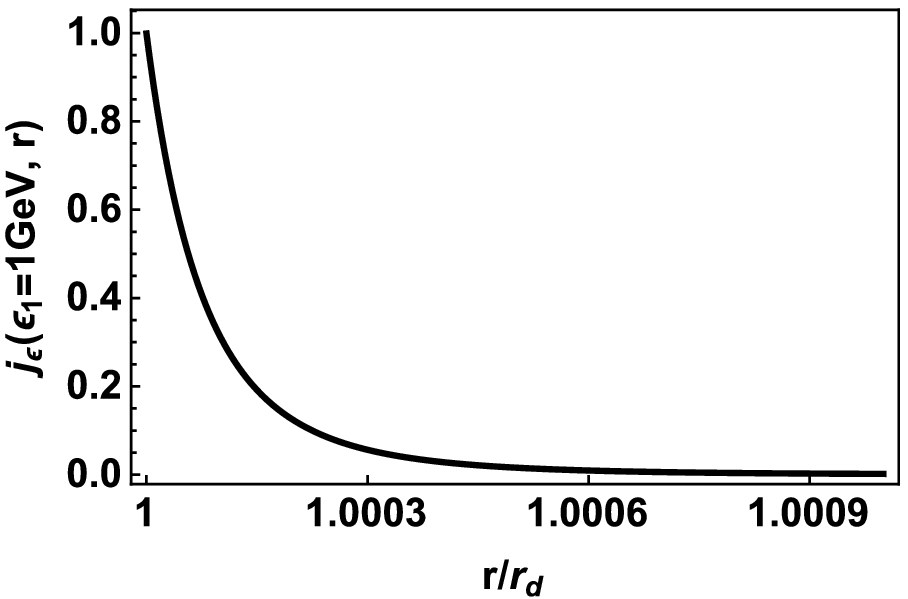}}
\centerline{\includegraphics[scale=0.8]{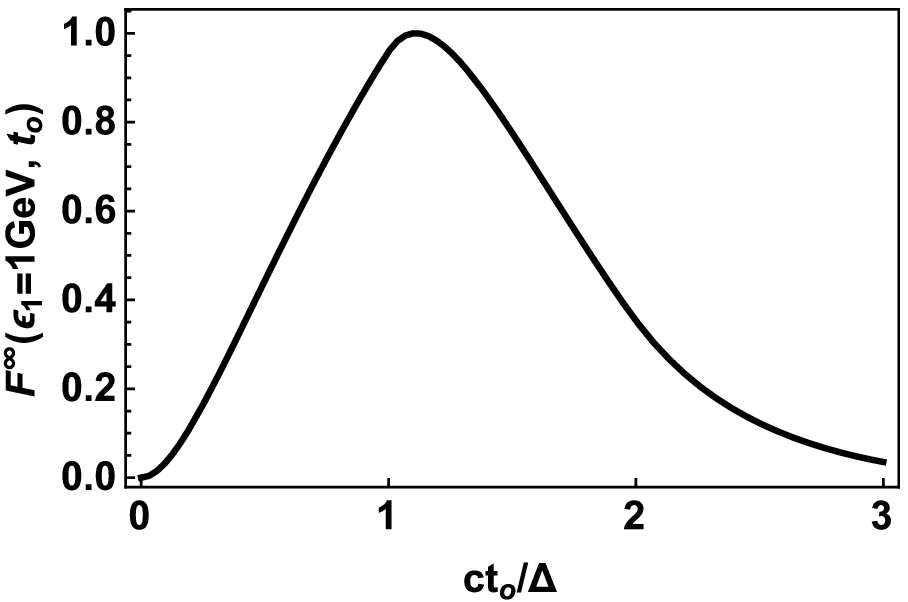}\includegraphics[scale=0.8]{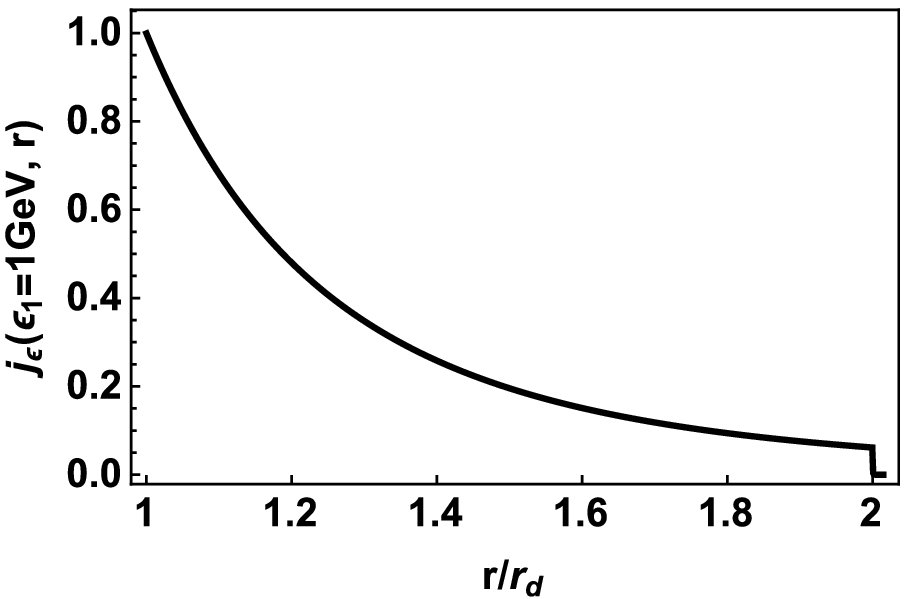}}
\caption{Sample light curves for non-delayed deceleration (upper panels) and delayed deceleration with $r_{dec}-r_d=\Delta/(1-\beta_\Gamma)$ (lower panels). The right panel in each case exhibits the (normalized) flux at a photon energy of 1 GeV 
measured by a distant observer as a function of  observer time, and the left panel the corresponding emissivity as a function of  
radius $r=ct$.  The two curves in the upper left panel correspond to different jet opening angles, as indicated.}
\label{fig-f4}
\end{figure*}

The evolution of the SED is  shown in figure \ref{fig-f3} in the case of immediate deceleration 
(left panel) and delayed deceleration 
with $r_{dec}=r_d+\Delta/(1-\beta_\Gamma)$ (right panel). 
The corresponding  lightcurves  are displayed in figure \ref{fig-f4} at a photon energy $\epsilon=1$ GeV.  
The right pannels exhibit the emissivity, and the left pannels the observed 
flux computed for these emissivities using Equation (\ref{ob-flux-curvt}).  The effect of the jet opening angle on the shape of the
lughtcurve is more prominent in the non-delayed case, as seen in the upper left panel of figure \ref{fig-f4};
in the delayed case this effect is essentially negligible.  These lightcurves are rather typical
in the regime where the deceleration time of the blob, $r_d/(c \alpha)$, is not much larger than its light crossing time 
$\Delta/(1-\beta_\Gamma)$.  In this regime, the duration of the flare measured by a distant observer viewing the 
source at an angle smaller than the opening angle of the flow and the overall shape of the lightcurve depend 
on details, as seen in figure \ref{fig-f4}.

\begin{figure*}
\centerline{\includegraphics[scale=1.0]{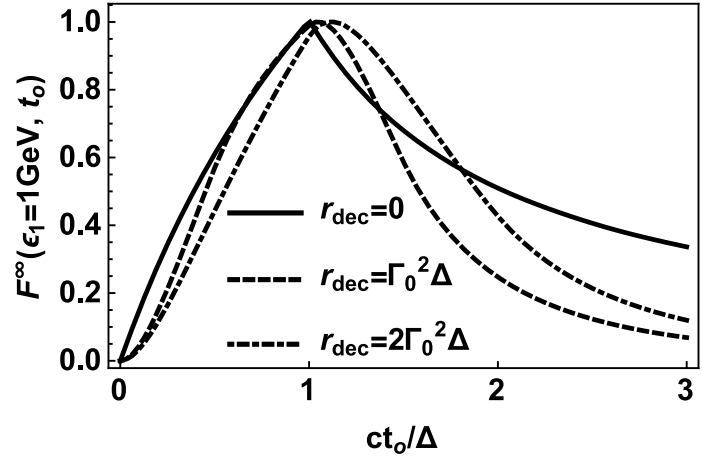}}
\caption{The effect of delayed deceleration on the shape of the light curve.  The jet opening angle in all cases shown is $\theta_j=\pi/2$.}
\label{fig-f5}
\end{figure*}

In the case of immediate deceleration the lightcurve is asymmetric, with the rise and decay times determined by the curvature 
of the emitting surface.  
For a sufficiently large Lorentz factor, such that the beaming angle is smaller than the opening angle of the flow (i.e., $\Gamma_0\theta_j>1$), these times, as measured by a distant observer, satisfy  $t_{rise}\sim t_{decay}\simeq r_d/\Gamma_0^2$. 
Those times might be comparable to the light crossing time of the blob if dissipation commences at a 
radius $r_d\simeq \Gamma_0^2\Delta$, as in the case shown in \ref{fig-f4} and \ref{fig-f5}.
In the case of  delayed deceleration the  rise and decay times are determined by the delay: 
$t_{rise}\simeq t_{decay}\simeq(r_{dec}-r_d)(1-\beta_\Gamma)$.  The lightcurve tends to be more symmetric in this case 
(see figs \ref{fig-f5} for a comparison).   The change in the emissivity during the coasting (pre-deceleration) phase, 
as seen in the lower right panel in figure \ref{fig-f4},  is due to the dependence of the intensity of target photons 
and the density of emitting electrons on $r$ ($j_{sc}\propto 
r^{-4}$ for the conical flow considered in the above  examples, with intensity profile $f_s=x^{-2}$ in Equation (\ref{Is-star})).   
This dependence may slightly affect the shape of the lightcurve.

The shape of the lightcurve displayed in the lower left panel of figure \ref{fig-f4} is in qualitative agreement 
with those computed in  Joshi et al. (2014), and is quite typical to observed gamma-ray flares.  The rough symmetry of the light curve is due to the assumed uniformity of the emissivity.  In cases where the acceleration of the electrons is confined to 
a small region inside the shell, e.g., acceleration in shock fronts, we anticipate a faster rise and a slower decay, particularly
at very high energies, at which the cooling time is much shorter than the light crossing time of the shell. 

\section{Discussion}

We considered the dynamics of a dissipative shell in the presence of a strong radiative drag.  Our analysis indicates
that for rapid dissipation substantial deceleration is anticipated in blazars on sub-parsec scales (and even 
parsec scales for sufficiently luminous sources), that should give rise to 
rapid, large amplitude variability owing to changes in the beaming factor.   
It is worth noting that even modest changes in the Lorentz factor can lead to large amplitude variations of the
emitted flux, owing to its sensitive dependence on the Doppler factor.  
In principle, this mechanism can produce flares with durations as short as 
the dynamical time of the central engine, at a very high efficiency.
The radiative drag exerted on the 
thermal electrons alone should lead to rapid deceleration at radii $r<4\times10^{17}\Gamma_0L_{s44}$ cm, 
where $\Gamma_0$ is the bulk Lorentz factor at the onset of dissipation and $L_s=10^{44}L_{s44}$ erg s$^{-1}$ is the luminosity 
of the ambient radiation field intercepted by the flow, provided the dissipation is rapid enough to keep the specific enthalpy large (that is, $h>>1$).  This is, for instance, the situation in internal shocks, as shown in section \ref{sec:shocks}, but may also occur in other cases, e.g., effective magnetic field dissipation in Poynting flux dominated jets (Sikora et al. 1996).
Scattering  of ambient photons by non-thermal electrons accelerated in 
situ will dominate the radiative force if the proper scale over which electrons are accelerated 
exceeds $ct^\prime_{c}/\xi_e$, where $t^\prime_c$ is the cooling time of the non-thermal electrons, as measured in 
the rest frame of the flow, and $\xi_e$ is the fraction of 
total energy carried by the non-thermal population.  

In general, the effect of radiative drag is expected to be more prominent in FSRQs than in BL Lacs.  Nonetheless,
Equation (\ref{alpha}) indicates that deceleration of dissipative shells may also be relevant to low luminosity sources,
provided that electrons can be effectively accelerated to the cooling cutoff.   This seems to be case in TeV blazars.  
It has been shown elsewhere (Levinson 2007) that if the TeV spectrum  emitted during strong flares extends to energies at which the 
pair production opacity exceeds unity deceleration should be effective.  

It is naively anticipated that the gamma-ray emission produced through IC scattering of ambient photons by 
the nonthermal electrons accelerated in the blob will be correlated with lower energy emission generated by synchrotron 
cooling of the same electrons.  However, synchrotron self-absorption may give rise to a strong suppression of 
the radio emission in cases where the deceleration occurs well below the radio core.    In that case,  the gamma
ray flare will either precede the ejection of a superluminal component, or not be accompanied by one at all, depending on the 
asymptotic bulk Lorentz factor and the synchrotron cooling time.    On the other hand, the onset of dissipation depends on the duty
cycle (the time interval between ejections of consecutive shells in the case of internal shocks), and is expected to occur over a range
of scales, even in an individual object.    Blobs that dissipate their energy at large enough radii will not experience strong deceleration.
As a consequence, a variety in the behavior of flares is expected, as indeed revealed by recent multi-waveband studies (e.g., Marscher et al. 2011).

We thank Noemie Globus for help.  
This research was supported by  a grant from the Israel Science Foundation no. 1277/13.

\appendix
\section{\label{sec:appA}Derivation of the source terms}
The general expression for the source term associated with Compton drag is (Phinney 1982; Sikora et al. 1996, Van Putten \& Levinson 2012):
\begin{equation}
S^\mu_c=-\sigma_T\int\frac{d^3p}{p^0}\int\frac{d^3k}{k^0}f_rf_e p_\nu k^\nu\left[k^\mu+\frac{(p_\nu k^\nu)p^\mu}{m_e^2}\right]
\end{equation}
Here $f_r$ and $f_e$ denote the distribution functions of the photons and electrons, respectively. 
In the rest frame of the blob the electron distribution $f_e$ is isotropic, and it is convenient to compute the source terms there, and
then to transform.    We have
\begin{equation}
p'_\nu k^{'\nu}=-p^{'0}k^{'0}+\vec{p'}\cdot\vec{k'}=-p^{'0}k^{'0}+p'k'\mu=-p^{'0}k^{'0}(1-\beta\mu)
\end{equation}
where $\mu$ is the cosine of the angle between the photon and electron directions, and $p^{\prime 0}/m_e=\gamma$, $p^{\prime i}/m_e=\gamma\beta^i$.
The zeroth component reads:
\begin{eqnarray}
S^{\prime 0}_c&=&-\sigma_T\int\frac{d^3p'}{p^{\prime 0}}\int\frac{d^3k'}{k^{\prime0}}f'_rf'_e p'_\nu k^{'\nu}\left[k^{'0}+\frac{(p'_\nu k^{'\nu})p^{'0}}{m_e^2}\right]\\
& =& \sigma_T\int d^3k'f'_r\int d^3p' f'_e k^{'0}\left[(1-\beta\mu)-\gamma^2(1-\beta\mu)^2\right]\\
& =&  \sigma_T\int d^3k' k^{'0} f'_r\int dp' p^{\prime 2}f'_e \int d\mu\left[(1-\beta\mu)-\gamma^2(1-\beta\mu)^2\right].\\
\end{eqnarray}
Now, $n_e'=\int d^3p' f_e=2\int dp' p^{\prime2} f_e$, and using $\int d\mu (1-\beta\mu)^2=2(1+\beta^2/3)$ one has
\begin{eqnarray}
S^{\prime0}_c=-\frac{4}{3}\sigma_T\int d^3k' k^{\prime 0} f'_r\int d^3p' f'_e\gamma^2\beta^2=-\frac{4}{3}\sigma_Tn'_e<\gamma^2\beta^2>u'_s.
\end{eqnarray}
The spatial component reads:
\begin{eqnarray}
S^{\prime i}_c&=&-\sigma_T\int\frac{d^3p'}{p^{\prime0}}\int\frac{d^3k'}{k^0}f'_rf'_e p'_\nu k^{'\nu}\left[k^{\prime i}+\frac{(p'_\nu k^{'\nu})p^{\prime i}}{m_e^2}\right]\\
& =& \sigma_T\int d^3k'f'_r\int d^3p' f'_e k^{\prime i}\left[(1-\beta\mu)-\gamma^2\beta(1-\beta\mu)^2\right]\\
& =&  \sigma_T\int d^3k' k^{'i} f'_r\int dp' p^{'2}f'_e (2+\frac{4}{3}\gamma^2\beta^2)=\sigma_Tn'_e<1+2\gamma^2\beta^2/3>T_r^{'0i},
\end{eqnarray}
where 
\begin{equation}
T_r^{'\mu\nu}=\int \frac{d^3k'}{k^{'0}}k^{'\mu} k^{'\nu} f'_r
\end{equation}
Now, in the black hole frame the radiation field is isotropic and we have $T^{00}=u_s$, $T^{0i}=0$, $T^{ij}=(u_s/3)\delta_{ij}$.  Then
\begin{equation}
T_r^{'0 i}=\Lambda^0_\mu\Lambda^{i}_\nu T^{\mu\nu}=-\frac{4}{3}\Gamma^2\beta_{\Gamma}^iu_s.
\end{equation}
Thus, we finally have for the source terms in the comoving frame:
\begin{eqnarray}
& & S^{\prime 0}_c=-\frac{4}{3}\sigma_Tn'_e<\gamma^2\beta^2>u_s\Gamma^2(1+\beta_\Gamma^2/3),\label{S0prime}\\
& & S^{\prime i}_c=-\frac{4}{3}\sigma_Tn'_e<1+2\gamma^2\beta^2/3>u_s\Gamma^2\beta_\Gamma^i. \label{Srprime}
\end{eqnarray}
To obtain the source term in the Lab frame we perform a Lorentz transformation,
\begin{eqnarray}
& & S_c^{0}=\Gamma S_c^{'0}+\Gamma\beta_\Gamma^iS'_{ci}=-\frac{4}{3}\sigma_Tn'_e\Gamma^3u_s[<\gamma^2\beta^2>(1+\beta_\Gamma^2)+\beta_\Gamma^2].\label{S_c^0-lab} \\
& & S_c^{i}=\Gamma\beta_\Gamma^{i} S_c^{'0}+\Gamma S_{c}^{\prime i}=-\frac{4}{3}\sigma_Tn'_e\Gamma^3\beta_\Gamma^iu_s[1+(5+\beta^2_\Gamma)<\gamma^2\beta^2/3>]
\label{S_c^i-lab}
\end{eqnarray}
 For cold electrons $<\gamma^2\beta^2>=0$, and $S_c^{\prime0}=0$, as required, since in the Thomson limit invoked here
 the scattering is fully elastic and there should be no energy loss in the rest frame 
 of the blob if the electrons are cold.   In the Lab frame we then 
recover the result $S^0_c=-\frac{4}{3}\sigma_Tn'_e\Gamma^3\beta^2_{\Gamma}u_s$.

\section{\label{app:Eq-mot}Flow equations}

In spherical coordinates, the radial expansion of a hydrodynamic shell is governed by the equations
\begin{eqnarray}
\partial_t(w\Gamma^2-p)+\frac{1}{r^2}\partial_r(r^2 w\Gamma^2\beta_\Gamma)=S_c^{0},\label{mot-radial0-app}\\
\partial_t(w\Gamma^2\beta_\Gamma)+  \frac{1}{r^2}\partial_r(r^2 w\Gamma^2\beta^2_\Gamma)+\partial_r p=S_c^{r},
\label{mot-radialr-app}
\end{eqnarray}
where $S_c^0$ and $S_c^r$ are source terms that account, respectively,  for energy and momentum  losses by the radiative drag. 
The above equations can be combined to yield
\begin{equation}
w\Gamma^2\frac{d\beta_\Gamma}{dt}+\beta_\Gamma\partial_tp+\partial_r p=S_c^r-\beta_\Gamma S_c^0,\label{Eq-mot-appB}
\end{equation}
in terms of the Lagrangian derivative $d/dt=\partial_t+\beta_\Gamma\partial_r$.

We suppose that at time $t$ the dissipative shell is contained between the radii $r_1(t)$ and $r_2(t) = r_1(t)+\Delta(t)$.   
For simplicity we suppose that the length of the shell $\Delta$ is constant in the black hole frame.     The velocity of the shell
is then uniform with $\beta_\Gamma(t)=dr_2/dt$.
Transforming to the coordinates $\tau(r,t)=t$, $\zeta(r,t)=r_2(t)-r$, we have 
\begin{eqnarray}
\partial_r=-\partial_\zeta,\\
\partial_t=\partial_\tau+\beta_\Gamma\partial_\zeta,\\
\frac{d}{dt}=\partial_t+\beta_\Gamma\partial_r=\partial_\tau.
\label{coor-trans-app}
\end{eqnarray}
 Using the relation $\beta_\Gamma d\beta_\Gamma=\Gamma^{-3} d\Gamma$, Equation (\ref{Eq-mot-appB}) gives
\begin{equation}
w\frac{d}{dt}\ln \Gamma+\frac{dp}{dt}-\Gamma^{-2}\partial_\zeta p=S_c^r-\beta_\Gamma S_c^0,\label{Eq-mot-newcoord}
\end{equation}
to order $O(\Gamma^{-2})$.  Equation (\ref{mot-radial0-app}) yields

\begin{equation}
\frac{d}{dt}(w\Gamma^2-p)+\frac{2}{r}w\Gamma^2\beta_\Gamma -\partial_\zeta p-w\partial_\zeta \ln\Gamma=S_c^0.
\end{equation}
To order  $O(\Delta/r_2)$ we have $2\beta_\Gamma/r=(2/r_2)dr_2/dt=r_2^{-2}dr_2^2/dt$, and the latter equation  reduces to 
\begin{equation}
\frac{1}{r_2^2}\frac{d}{dt}(r_2^2w\Gamma^2)-\frac{d}{dt}p-\partial_\zeta p-w\partial_\zeta \ln\Gamma=S_c^0.
\end{equation}

\break 

\end{document}